\documentclass[10pt, conference, compsocconf]{IEEEtran}
\usepackage{cite}

\ifCLASSINFOpdf
\usepackage[pdftex]{graphicx}
\else
\fi
\usepackage{amssymb}
\usepackage{url}

\usepackage{textcomp}
\usepackage{cleveref}
\usepackage{listings}
\lstdefinestyle{lua}{language={[5.0]Lua},basicstyle=\footnotesize\ttfamily,frame=lines,captionpos=b}
\usepackage{wrapfig}

\usepackage[usenames,dvipsnames]{xcolor}
\usepackage{ifthen}

\usepackage{tikz}
\usepackage{textcomp}
\usepackage{lipsum}
\newcommand\copyrighttext{%
  \footnotesize 978-1-5090-6611-7/17 \$31.00 \textcopyright2017 IEEE.\\
    Personal use of this material is permitted.
    Permission from IEEE must be obtained for all other uses,
    in any current or future media, including reprinting/republishing this
    material for advertising or promotional purposes, creating new collective
    works, for resale or redistribution to servers or
    lists, or reuse of any copyrighted component of this work in other works.\\
    Pre-print version. For the final published paper, refer to
    DOI: \url{http://dx.doi.org/10.1109/CCGRID.2017.129}}
\newcommand\copyrightnotice{%
\begin{tikzpicture}[remember picture,overlay]
\node[anchor=south,yshift=10pt] at (current page.south) {\fbox{\parbox{\dimexpr\textwidth-\fboxsep-\fboxrule\relax}{\copyrighttext}}};
\end{tikzpicture}%
}

\newboolean{showcomments}
\setboolean{showcomments}{false}
\ifthenelse{\boolean{showcomments}}
{ \newcommand{\mynote}[3]{
    \fbox{\bfseries\sffamily\scriptsize#1}
    {\small$\blacktriangleright$\textsf{\emph{\color{#3}{#2}}}$\blacktriangleleft$}}}
{ \newcommand{\mynote}[3]{}}

\begin{document}
%
\title{A lightweight MapReduce framework for secure processing with SGX}


%


%
\author{\IEEEauthorblockN{Rafael Pires\IEEEauthorrefmark{1},
Daniel Gavril\IEEEauthorrefmark{2},
Pascal Felber\IEEEauthorrefmark{1}, 
Emanuel Onica\IEEEauthorrefmark{2} and
Marcelo Pasin\IEEEauthorrefmark{1}}
\IEEEauthorblockA{\IEEEauthorrefmark{1}University of Neuch{\^a}tel, Switzerland;\IEEEauthorrefmark{2}Alexandru Ioan Cuza University of Ia\c{s}i, Romania}
}

\maketitle
\copyrightnotice

\begin{abstract}
MapReduce is a programming model used extensively for parallel data processing in distributed environments. 
A wide range of algorithms were implemented using MapReduce, from simple tasks like sorting and searching up to complex clustering and machine learning operations.
Many of these implementations are part of services externalized to cloud infrastructures. 
Over the past years, however, many concerns have been raised regarding the security guarantees offered in such environments.
Some solutions relying on cryptography were proposed for countering threats but these typically imply a high computational overhead.
Intel, the largest manufacturer of commodity CPUs, recently introduced SGX (software guard extensions), a set of hardware instructions that support execution of code in an isolated secure environment.
In this paper, we explore the use of Intel SGX for providing privacy guarantees for MapReduce operations, and based on our evaluation we conclude that it represents a viable alternative to a cryptographic mechanism.
We present results based on the widely used k-means clustering algorithm, but our implementation can be generalized to other applications that can be expressed using MapReduce model.
\end{abstract}


%
\IEEEpeerreviewmaketitle

\section{Introduction}
\label{sec:intro}

Since it was officially adopted by Google \cite{Dean:2008} in 2008, the MapReduce programming model consistently gained ground as a viable solution for assuring the necessary scalability of distributed data processing. 
The generic model, composed of the two main \emph{map} and \emph{reduce} functions, was widely used to implement applications that can leverage parallel task processing.
In this generic model, the data to be processed is typically located over a series of mapper nodes, which apply in parallel a \emph{map} function responsible for mapping individual data items to a finite set of predefined keys. 
The output is redistributed in a shuffle step based on the key mappings to reducer nodes, which execute in parallel a \emph{reduce} function for processing each set of data items corresponding to a certain key.
This model can be used in processing tasks that range from simple operations that can be composed like counting, sorting, and searching data to more complex algorithms like cross-correlation or page rank. 
MapReduce was adapted in different ways to fit this wide diversity of scenarios, as well as different deployment environments.

Particular settings where the deployment of MapReduce encountered difficulties are services that require security guarantees.
For cost savings purposes, data analysis and processing services are more and more often deployed to public cloud infrastructures, which do not offer strong security guarantees and hence make sensitive data prone to privacy and integrity risks.
If such services rely on a MapReduce-based implementation, it is of high importance to find a way to adapt the programming model in a manner that provides the required security assurance to the system.
As we refer in Section~\ref{sec:rw}, some solutions use complex cryptographic techniques for protecting the privacy of the data being processed in the \emph{map} and \emph{reduce} functions. 
Such techniques are typically complex: they range from secret sharing to homomorphic encryption, depending on the exact operations that need to be executed over the sensitive data in each use case.
In particular the choice of the encryption is largely determined by the task to be executed in the reducers.
Therefore, a common limitation is that the customized data encryption schemes used often have limited expressiveness and are applicable to only a small class of applications. 
Another drawback generally encountered is the high performance overhead of complex encryption schemes, homomorphic encryption in particular being notorious for its lack of practicality.
Finally such solutions address mostly the privacy of the data, but much less often the privacy or the integrity of the code itself, which is a key concern when externalizing a service to a public cloud.

A different approach to security is to rely on trusted execution environments (TEE) supported in hardware. 
A TEE provides an isolated space for executing code where confidentiality and integrity are assured.
\emph{Software guard extensions} (SGX) is a powerful TEE included in the Intel's commodity processors starting with the Skylake generation in late 2015.
The SGX-capable processors provide the possibility of running code in \emph{enclaves} that are isolated from the other memory used by system processes. 
This facilitates the provision of privacy and integrity for proprietary algorithm implementations and for sensitive data.
The code designated to run in the enclave space is not accessible from outside. 
Sensitive data can be protected using state-of-the-art encryption algorithms while outside the enclave and decrypted only inside where is processed efficiently in plaintext form---but shielded from the rest of the system.

In our work we propose a self-contained framework for securing MapReduce that leverages the benefits of SGX's trusted execution environment.
Our system architecture combines a lightweight virtual machine based on the Lua language, a MapReduce library, and a publish/subscribe service for communication between the client and worker nodes.
Unlike cryptographic solutions, our approach is independent of the particular characteristics of the \emph{map} and \emph{reduce} functions and can hence be used for any problem parallelizable with MapReduce.
The specific code to be executed in the MapReduce service can be integrated in simple scripts, which are run privately and in isolation using SGX over data decrypted only inside the enclave.
Our focus is on providing a flexible framework for securely running MapReduce applications that can be easily implemented and deployed.
The basic \emph{word count} MapReduce example, for counting the number of occurrences of different word in a given text, can be implemented in our framework with less than 30 lines of code (LOC). 
We present in this paper an evaluation of our approach using the widely used k-means clustering algorithm, showing that the overhead incurred is minimal and that our solution is applicable to other use cases.

Our paper is structured as follows. 
We overview related work in Section~\ref{sec:rw}, including other approaches on securing MapReduce based services. 
We then present additional details on SGX and k-means in Section~\ref{sec:preliminaries}, and we describe the architecture of our solution in Section~\ref{sec:arch}.
Section~\ref{sec:eval} covers the evaluation of the k-means use case by comparing our SGX-secured architecture with a basic unsecured implementation.
We finally conclude in Section~\ref{sec:conclusion}.

\section{Related Work}
\label{sec:rw}

VC3 \cite{Schuster:Security:2015} is an early proposal of a secure framework for MapReduce services using an SGX emulator instead of real SGX-enabled processors.
The general idea behind the framework design is relatively close to our proposal, the user having the possibility to write its own \emph{map} and \emph{reduce} functions that can execute in SGX-enabled machines. 
The user code has to be written in C++, which can make the implementations prone to potential faults like illegal memory accesses. 
In this respect the authors provide an optional compiler through which the programmers can enforce self-integrity invariants for memory regions. 
The main advantage in our proposed framework, which we more accurately evaluate effectively on SGX-enabled processors, is the additional flexibility and ease-of-use of the high-level Lua-based programming environment.
The MapReduce implementation in our case is enclosed in scripts, which are executed by a standard Lua interpreter that is less prone to faults, easier to maintain, and safer thanks to its small code base.
The VC3 solution is further extended in \cite{Ohrimenko:2015}, which focuses on security issues generated by traffic analysis attacks on the exchanges between mapper and reducer nodes. 
We do not consider such attacks in our current work, but we believe that the analysis and the solutions proposed are also applicable to our system.

$M^{2}R$ \cite{M2R} is another proposal of a secure framework for MapReduce, which takes a more general approach with a generic design that can be implemented on any trusted execution environment.
The authors refer to SGX-enabled processors as one potential target, but the evaluation is conducted as in case of VC3 above on a TEE simulation using a trusted Xen-4.4.3 kernel-based hypervisor. 
As in \cite{Ohrimenko:2015}, $M^{2}R$ specifically focuses on attacks exploiting the leakage between mappers and reducers. 
$M^{2}R$'s generic design formalizes the execution of the MapReduce architecture in four \emph{trusted code base} (TCB) components: \emph{mapT}, \emph{reduceT}, \emph{mixT}, and \emph{groupT}, with the security of the latter two being particularly critical as they are meant to contain the mappers-to-reducers shuffle implementation.
Evaluation using the HiBench MapReduce benchmark suite \cite{Huang:2010} yields the interesting observation that \emph{k-means clustering} (which we also consider in our tests) and \emph{grep}, two computationally intensive cases, do not leak extra information in the baseline implementation without the shuffle phase's extra security.
Therefore, in such situations, adding additional protection between mappers and reducers will produce unnecessary overheads.

In \cite{Dolev:2016} the authors propose a solution that leverages Shamir secret sharing \cite{Shamir:1979} for executing queries over an outsourced database using the MapReduce model. 
Each field in a database record is split in secret shares that are constructed for each letter in the field value. 
Each letter in the alphabet is associated with a unary vector of 26 bits, with one index bit set and the rest being zeroes. 
The secret shares are obtained for each bit by applying 26 polynomials of identical degree. 
Shares of numerical values, as well as the privacy preserving queries submitted by users, are represented in the same fashion.
Mappers and reducers execute specific operations depending on the queries on these shares for obtaining the result. 
The individual computations seem rather simple and efficient, since many of the initial bits are zeroes, but the total number of values obtained in the encoding (26 per letter) still inflicts a high overhead.
Furthermore, the technique is limited to four query types: count, search, equijoin, and range. 

SecureMR~\cite{SecureMR} focuses on MapReduce integrity. 
In this purpose some of the MapReduce tasks are replicated and assigned to different mappers and reducers.
The MapReduce architecture is enhanced with several secure components: a secure manager and a secure scheduler for task duplication and assignment, a secure task executor for DoS and replay attacks prevention, a secure committer for consistency checking of mappers intermediate results, and a secure verifier that detects attacks based on results inconsistencies.
The solution takes in account neither the privacy of data nor of the code.

Airavat~\cite{Roy:2010} proposes to secure MapReduce using differential privacy, which is also combined with access control policies provided by SELinux in the proposed implementation. 
The mappers can be both trusted and untrusted in the proposed architecture, the approach relying on adding noise in this phase in order to minimize leaks of sensitive data in the \emph{map} function output. 
Among the limitations of the approach, one can mention that reducers should be always trusted and that there is no possibility to chain multiple MapReduce cycles on the same input data.
Moreover, the accuracy of the results depends on parameters that should be carefully tweaked for obtaining a sound output.

Tagged MapReduce~\cite{TaggedMR} considers the execution of MapReduce computations over a hybrid cloud, where a part of the infrastructure is public and untrusted and a part is private and trusted. 
The sensitive data can be processed securely only in the private part and is identified through tags added to the tuples. 
For this purpose, the system architecture involves a scheduler that assigns tasks to workers and control the flow based on tags.
The programmer can implement in \emph{map} and \emph{reduce} functions specific policies that dictate how data sensitivity changes during the processing.
The solution does not address a case where only a public untrusted cloud is sufficient for deployment.

\section{Preliminaries}
\label{sec:preliminaries}

The MapReduce model includes three main sequential phases in processing the input data, which we also briefly described in Section~\ref{sec:intro}. 
In the first phase the mapper nodes independently execute a \emph{map} function in which each individual item in the distributed input data is mapped to a key.
In the second phase the (key, value) tuples obtained after mapping are shuffled according to their keys and redistributed to reducer nodes, such that all tuples corresponding to the same key arrive on the same node.
Optionally, in this phase multiple tuples with the same key can be aggregated by a combiner function in order to optimize the redistribution. 
Finally, in the third phase reducer nodes process independently in a local \emph{reduce} function all data associated to a key and output the result of their computation.
In our work, we are concerned with assuring the privacy and integrity of the data processed within the \emph{map} and \emph{reduce} functions and also of the code of these functions, while also offering to the programmer an accessible lightweight environment of implementing various use-cases.
In the current solution we do not consider any attacks targeting the in-between shuffle phase like traffic analysis, leaving this for future work. 
In the framework we propose data is encrypted outside of the two main \emph{map} and \emph{reduce} functions, which are executed securely in SGX enclaves. 

In our architecture we use the publish/subscribe (pub/sub) communication model~\cite{CS:03}, one of the most effective ways of disseminating information in distributed systems.
In the generic model, publisher nodes submit data to a pub/sub routing service as publications formed of a header describing the data and a payload containing the effective data.
The pub/sub routing service matches the publications header with subscriptions registered by subscriber nodes and further routes the matching publications towards their destinations.  
Our solution is based on a previous Secure Content-Based Routing (SCBR) implementation~\cite{Pires:2016} that evaluated the pub/sub communication model in an SGX secured environment. 
In brief, in SCBR the matching and routing of publications is determined inside secure SGX enclaves provided by the pub/sub service, which is typically deployed in a public cloud.
All subscriptions and publication messages are encrypted using a symmetric cypher while outside the SGX enclaves. 
The subscriptions and publication headers are decrypted inside the enclave, where subscriptions are stored. Then, the service routes the publication payloads (encrypted with a different key) to matching subscribers.
For simplicity we leave out the details related to the key provisioning, which can be consulted in~\cite{Pires:2016}. 

We adapt the SCBR pub/sub implementation to a MapReduce service by designing a pub/sub communication protocol between the nodes involved in the distributed data processing. 
Worker nodes will act as subscribers registering queries to find out about new MapReduce job openings, and also as publishers submitting their desired job details. 
The client of the MapReduce service, which is also the data owner will also act as both subscriber and publisher, registering subscriptions for job details and publishing announcements on new MapReduce jobs to be executed. 
The client will also coordinate the routing from mapping to reducers using the pub/sub system, and will finally publish the code and data to the registered workers and obtain the results.
We detail our solution architecture, describing the exact steps executed in the pub/sub based communication protocol in~\ref{sec:arch}. 

\begin{figure}
\centering
\includegraphics[scale=0.35]{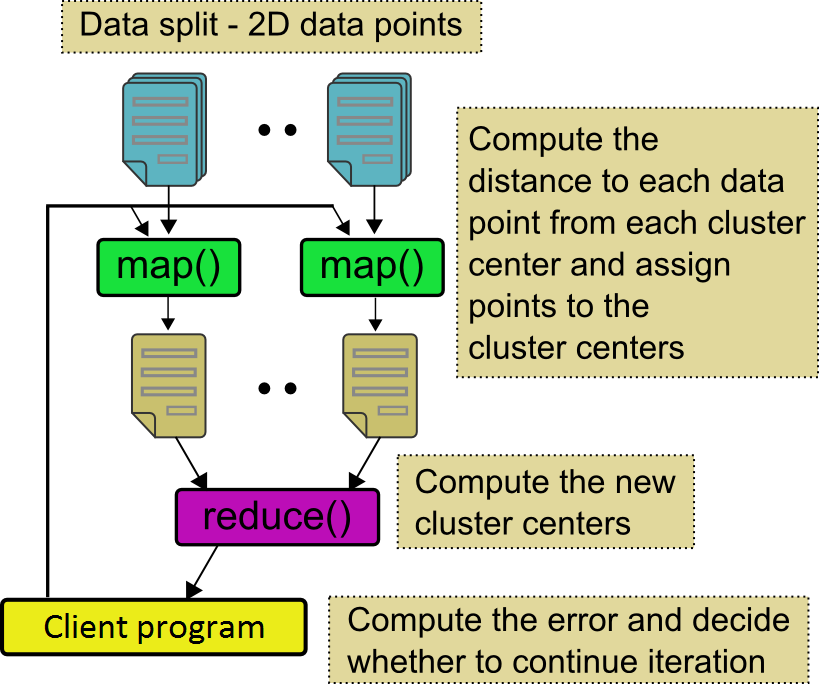}
\caption{\label{fig:mapreduce}Flow of K-means with MapReduce.}
\end{figure}

For testing our solution we consider a use case where the target of the MapReduce service is to classify data using the k-means clustering method. 
K-means is an unsupervised learning algorithm widely used and adapted for data classification purposes since its first mention in~\cite{macqueen1967}, which operates as following:
(i) a certain $k$ number of clusters is fixed \emph{a priori} and $k$ corresponding centers are defined for each one of them; 
(ii) data items are iterated and assigned to the nearest cluster center (typically through Euclidean distance); and
(iii) every cluster center is recomputed as the centroid of the assigned data items (the mean of those points). 
Steps (ii) and (iii) repeat until a termination criteria is reached 
(e.g. the sum of distances between the old and the new cluster centers is below a given threshold).
For implementing k-means in the MapReduce model we put (ii) in the \emph{map} function and (iii) in the \emph{reduce} function as displayed in~\Cref{fig:mapreduce}.

For providing an easily programmable and efficiently maintainable environment our framework offers the possibility to write the \emph{map} and \emph{reduce} code in simple Lua scripts, which are run in the enclave using a Lua interpreter, currently 
in version 5.3.2.
Lua~\cite{Lua2007} is a lightweight multi-paradigm programming language.
The language API provides the possibility to call Lua functions from C/C++ code. 
Using Lua is particularly interesting in an SGX secured environment due to the reduced overhead, which fits with the enclave restrictions.
We discuss in further detail the evaluation of the k-means use case with our Lua implementation in Section~\ref{sec:eval}.

\section{Solution Architecture}
\label{sec:arch}

\begin{figure}
\centering
\includegraphics[width=0.4\textwidth]{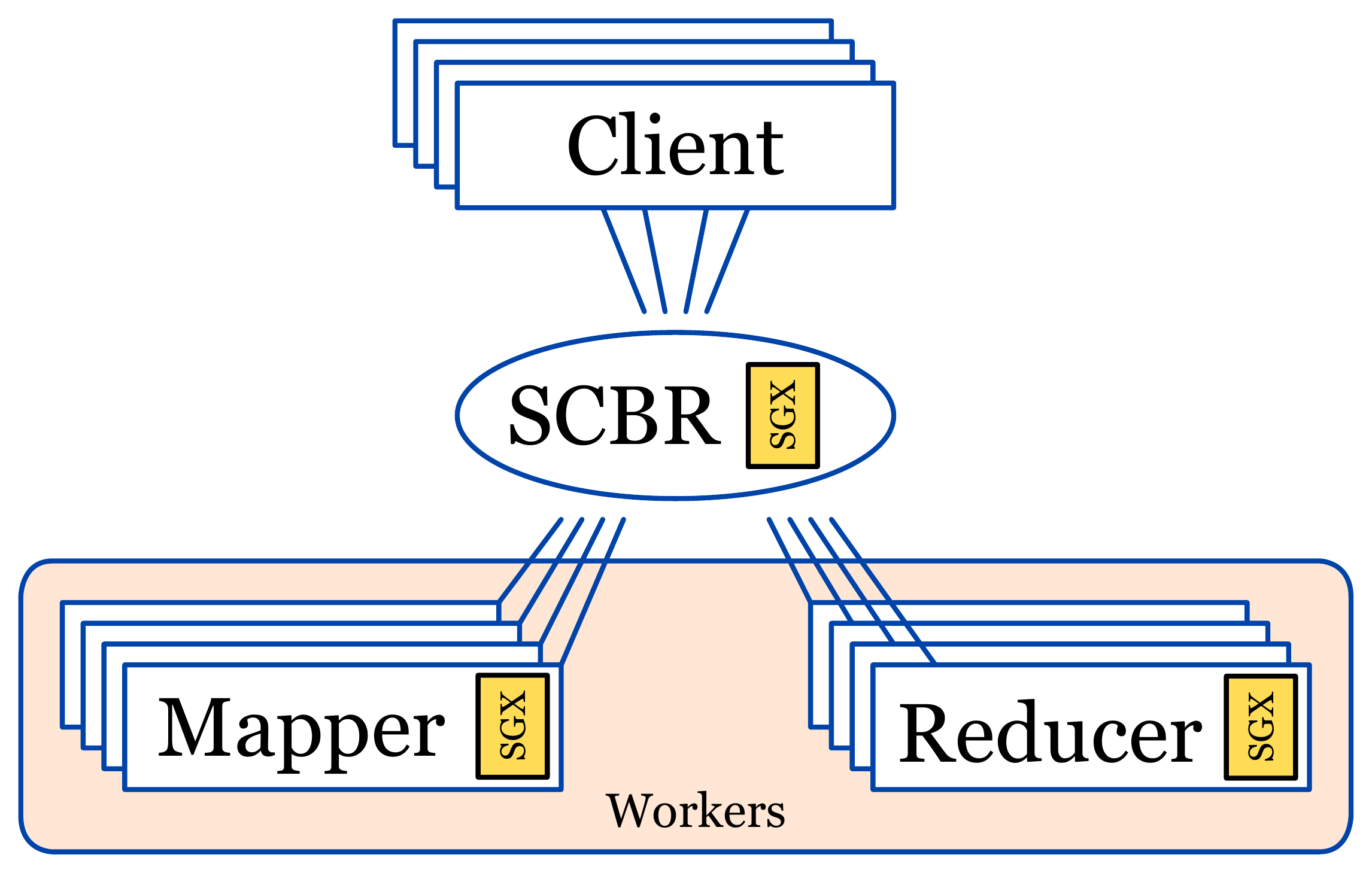}
\caption{\label{fig:entities}Entities.}
\end{figure}

In \Cref{fig:entities} we display the entities composing our solution architecture: clients, router (SCBR based pub/sub engine) and workers, which can assume the role either of a mapper or
a reducer. 
Clients provide the code to be executed, the data to be processed and gather the results after completion.
All communication channels use the 0MQ \cite{hintjens2013zeromq} message passing library, having a central point in the SCBR pub/sub engine, which we adapted from previous work~\cite{Pires:2016} as briefly described in Section~\ref{sec:preliminaries}.

The SCBR engine is responsible for securely storing subscriptions that contain the conditions under which each message should be forwarded to the corresponding interested party.
Subscriptions, used in an initiation protocol for the MapReduce processing as described further below, are stored within the SGX protected memory area and every match processing is done inside enclaves. 
Data is always encrypted whenever outside of the enclaves.
Although such a centralized approach is not suitable to large-scale data processing, it is arguably useful for modest quantities of highly sensitive data that could be, for instance, partitioned from higher amounts of non-sensitive data.
Nevertheless, it has been demonstrated~\cite{StreamHub2013,StreamHub2014} that it is possible to elastically scale a pub/sub engine by specializing its functional steps into replicable operators. 
That could dramatically improve the network performance of such a centralized approach. 

\begin{figure}
\centering
\includegraphics[scale=0.45]{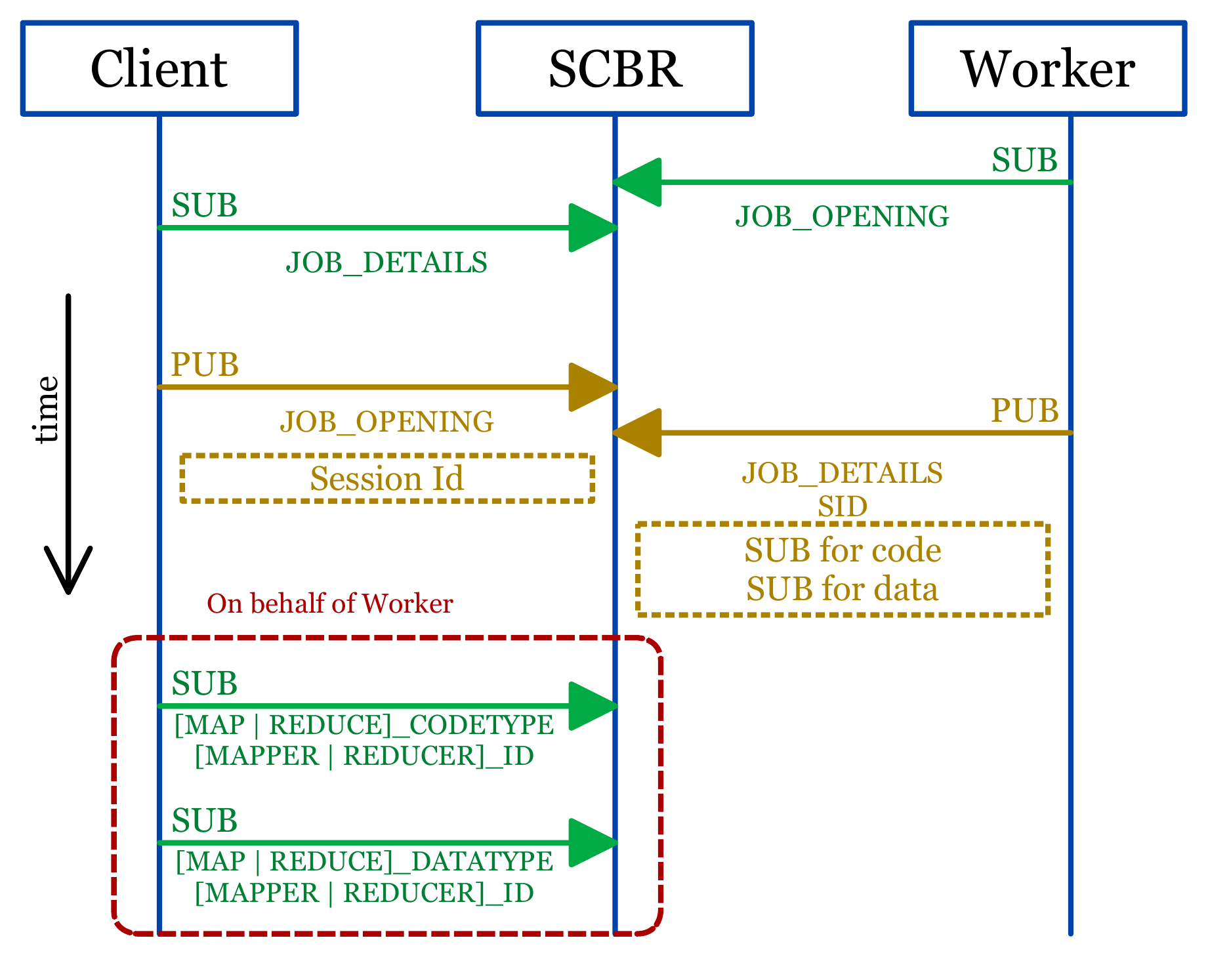}
\caption{\label{fig:initprotocol}Session establishment protocol.}
\vspace{-10pt}
\end{figure}

The MapReduce processing is started following an initial message exchange displayed in \Cref{fig:initprotocol}.
Worker nodes register their availability for MapReduce job openings through \texttt{JOB\_OPENING} subscriptions.
The client registers his interest in the details of the \emph{map} and \emph{reduce} tasks that the workers are capable of doing through the \texttt{JOB\_DETAILS} subscription.
When the client has a new job to execute it advertises it through a \texttt{JOB\_OPENING} publication, which is received by workers that previously registered as available.
The workers submit then their desired details regarding the job through a \texttt{JOB\_DETAILS} publication by sending in the message payload their further subscriptions for code and data (particular to the role they choose: mapper or reducer).
At the end of this negotiation, if the client decides to hire a worker, it registers on SCBR the received subscriptions for code and data on the worker's behalf. 
By doing so, the client established the MapReduce chain and keeps track of how many workers it has hired and of which kind (mappers or reducers).


\begin{figure}
\centering
\includegraphics[scale=0.45]{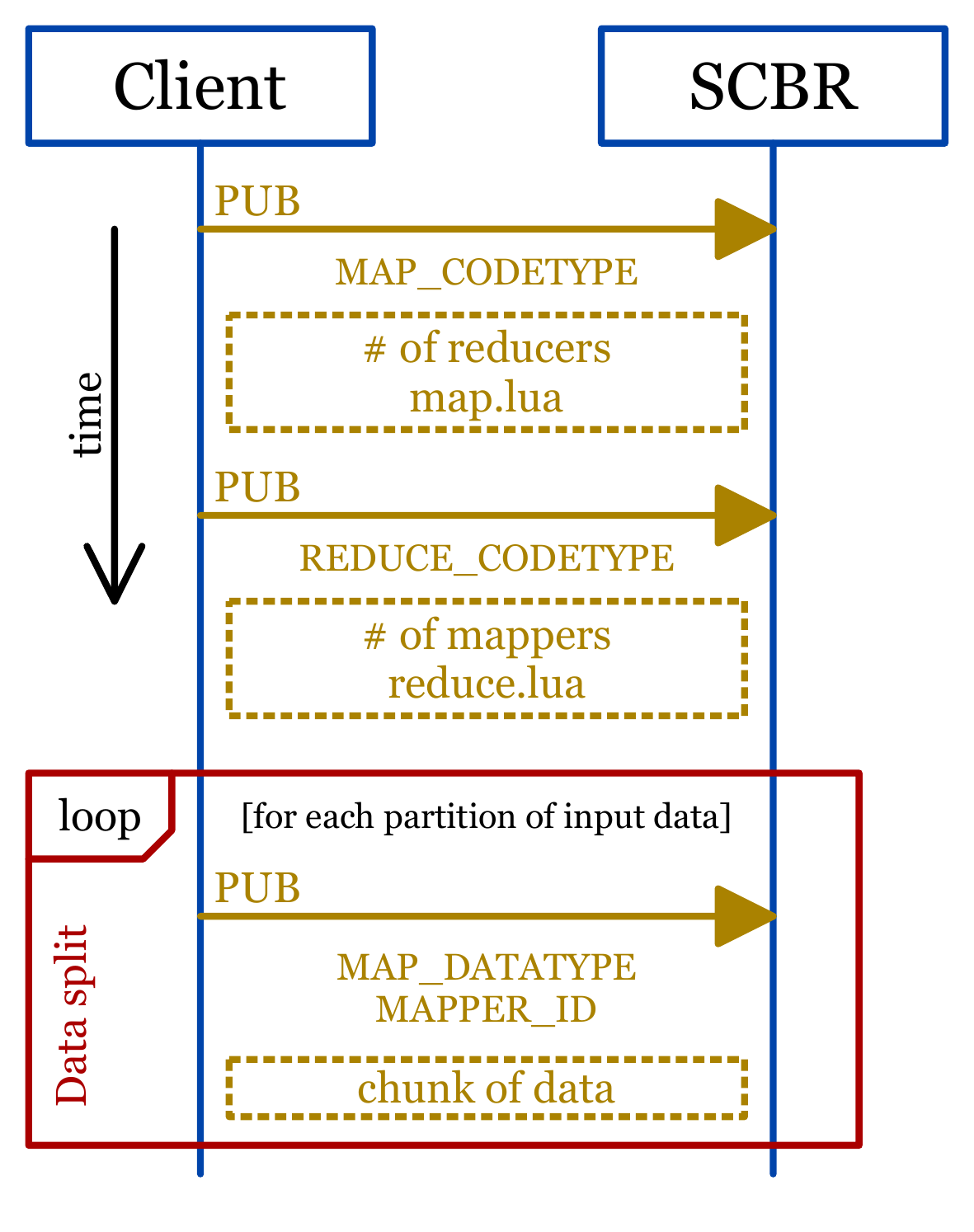}
\caption{\label{fig:provisioning}Provisioning of code and data.}
\vspace{-10pt}
\end{figure}

The provisioning of code and data is shown in \Cref{fig:provisioning}.
The client also includes the number of reducers along with the Lua scripts in case of type \texttt{MAP\_CODETYPE}, or the number of mappers in case of type \texttt{REDUCE\_CODETYPE}. 
The purpose is to make the workers aware of how many messages signaling the end of the stream they have to wait before considering the work done. 
This is important because the reduce phase can only start once all data for a given key is routed to the intended worker. 
Additionally, the number of reducers received by a mapper is used in a hash function that takes as argument the mapped key and returns the indication of which reducer it has to be forwarded to.
After sending the code, data is split by the client among the mappers, line by line. 
The destination id is included in the header of the \texttt{MAP\_DATATYPE} publication.

Workers decrypt the received code and store inside the enclaves waiting to start the execution.
When data arrives, mappers perform the processing and each one of the resultant set of key-value pairs is forwarded to the proper reducer. 
Its id is obtained after providing the key and number of reducers as arguments to the hash function that comes along with the code of the mapper. 
The shuffling phase is hence conducted by the mappers. 
In order to forward data to the following step, all the Lua script has to do is to call a special function called \emph{push(key,value)}, and the framework handles all the communication aspects of forwarding the data.

Regarding the execution environment, since SGX code must be previously signed and no dynamic linking is possible after the enclave creation, we ported a Lua \cite{Lua2007} virtual machine to run in protected area. 
Porting legacy code to SGX means that every system call or input/output instruction have to be dealt with, since these are not allowed inside enclaves.
Workers contain a Lua interpreter that runs inside the enclaves, which loads the scripts that contain the processing phases of \emph{map} and \emph{reduce}.
This setting provides a very accessible programming environment, which can be easily used and efficiently maintained.
We provide below a basic example.

\begin{lstlisting}[style=lua,caption={Map code in Lua for word count},label={lst:lua:map}]
local json = require "json"

function hash(key,rcount)
    return string.byte(key,1) % rcount
end

function combine(key,value)
    local clist = json:decode(value)
    local sum = 0
    for k,v in pairs(clist) do
        sum = sum + v
    end
    push(key,sum)
end

function map(key,value)
    for word in value:gmatch("%w+") do
        push(word,1)
    end
end
\end{lstlisting}

Listing \ref{lst:lua:map} shows the sample code of a mapper of a word count application.
As it can be noticed, data is encoded in JavaScript Object Notation (\emph{json}).
For convenience, the json Lua parser is pre-loaded in the enclave of worker nodes, and is accessible through the \emph{require} command.
The script can contain as many helper functions as desired. 
The following special functions, however, are called by the framework:

\vskip 2mm
\begin{tabular}{p{3cm}p{45mm}}
   \emph{map(key, value)} &
    Contains the functional implementation of mapper. \\[3pt]
   \emph{combine(key, value)} &
    [Optional] Post-processing on values grouped by keys. \\[3pt]
   \emph{hash(key,rcount)} &
    Returns the Reducer id that is supposed to receive a given \texttt{key} 
    considering that there are \texttt{rcount} Reducers in total. \\
\end{tabular}
\vskip 2mm

Likewise, listing \ref{lst:lua:reduce} shows a sample code for the reduce step that contains a single special function:

\vskip 2mm
\begin{tabular}{p{3cm}p{45mm}}
   \emph{reduce(key, value)} &
    Contains the functional implementation of reducer. \\[3pt]
\end{tabular}

\begin{lstlisting}[style=lua,caption={Reduce code in Lua for word count},label={lst:lua:reduce}]
local json = require "json"

function reduce( key, value )
    local clist = json:decode(value)
    local sum = 0
    for k,v in pairs(clist) do
        sum = sum + v
    end
    push(key,sum)
end
\end{lstlisting}

\section{Evaluation}
\label{sec:eval}

We construct our evaluation using the k-means algorithm, briefly described in Section~\ref{sec:preliminaries}, a popular use-case for cluster analysis in data mining and machine learning.
In our implementation, we partitioned \emph{n} randomly generated bi-dimensional coordinates into \emph{k} clusters. In the \emph{map} step, we assign each observation to the nearest center by computing and comparing their distances ($n\times k$ operations). 
Then, we update the centers to become the centroid of assigned coordinates ($n$ operations) in the \emph{reduce} step.
This process repeats until the average distance between the most recent calculated centroids and the ones from the previous iteration is less than a threshold. 

We conducted all the experiments using two SGX capable machines, both with processor Intel i7-6700 64bits, clock of 3.4GHz, 8MB cache, 4 cores, 8 threads, with 8GB of installed memory and SSD of 256GB. 
In terms of software, we used the Intel SGX SDK 1.7.100 over Ubuntu 14.04.1, kernel 4.2.0-42.
Unless mentioned otherwise, messages were all encrypted with AES-CTR~\cite{Daemen:2002} with key and input vector both of 128 bits and were decrypted only inside the enclaves. 
The process placement was made as follows: 
Machine 1: The Client, entity responsible for providing code and data, 8 mappers and 5 reducers. 
Machine 2: 8 mappers and 5 reducers.
The number of mappers was chosen to be twice as much as the number of cores in each machine to take advantage of parallelism, while the number of reducers was set to be a divisor of input data size to stimulate an even distribution of work among them.
To illustrate how small our final code-base is, \Cref{tab:codesize} shows the memory section sizes of executables and shared libraries that are loaded into enclaves.

\begin{table}[h!]
    \centering
    \begin{tabular}{|c|c|c|c|c|}
    \hline
                &text       &data       &bss    &sum     \\
    \hline
client          &379,968    &26,672     &376    &407,016 \\
    \hline
worker          &288,557    &26,312     &768    &315,637 \\
    \hline
worker enclave  &679,175    &60,004     &83,584 &822,763 \\
    \hline
scbr            &277,468    &14,808     &72     &292,348 \\
    \hline
scbr enclave    &278,967    &7,456      &80,608 &367,031 \\
    \hline
    \end{tabular}
    \caption{\label{tab:codesize}Binaries size in bytes}
\vspace{-10pt}
\end{table}

\begin{figure}
\centering
\includegraphics[width=0.43\textwidth]{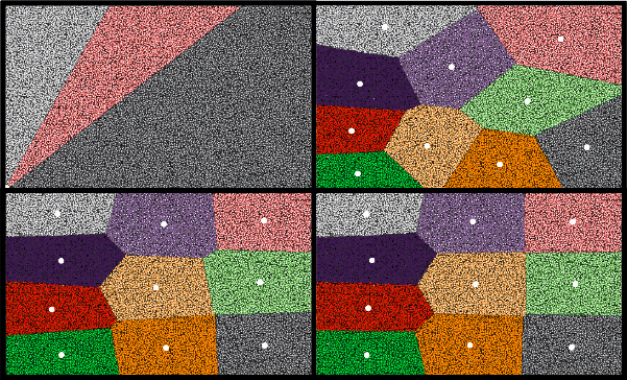}
\caption{\label{fig:kmeans}K-means convergence. Iterations 0 (top left), 19 (top right), 38 (bottom left) and 76 (bottom right)}
\end{figure}

\begin{figure}
\centering
\includegraphics[width=0.5\textwidth]{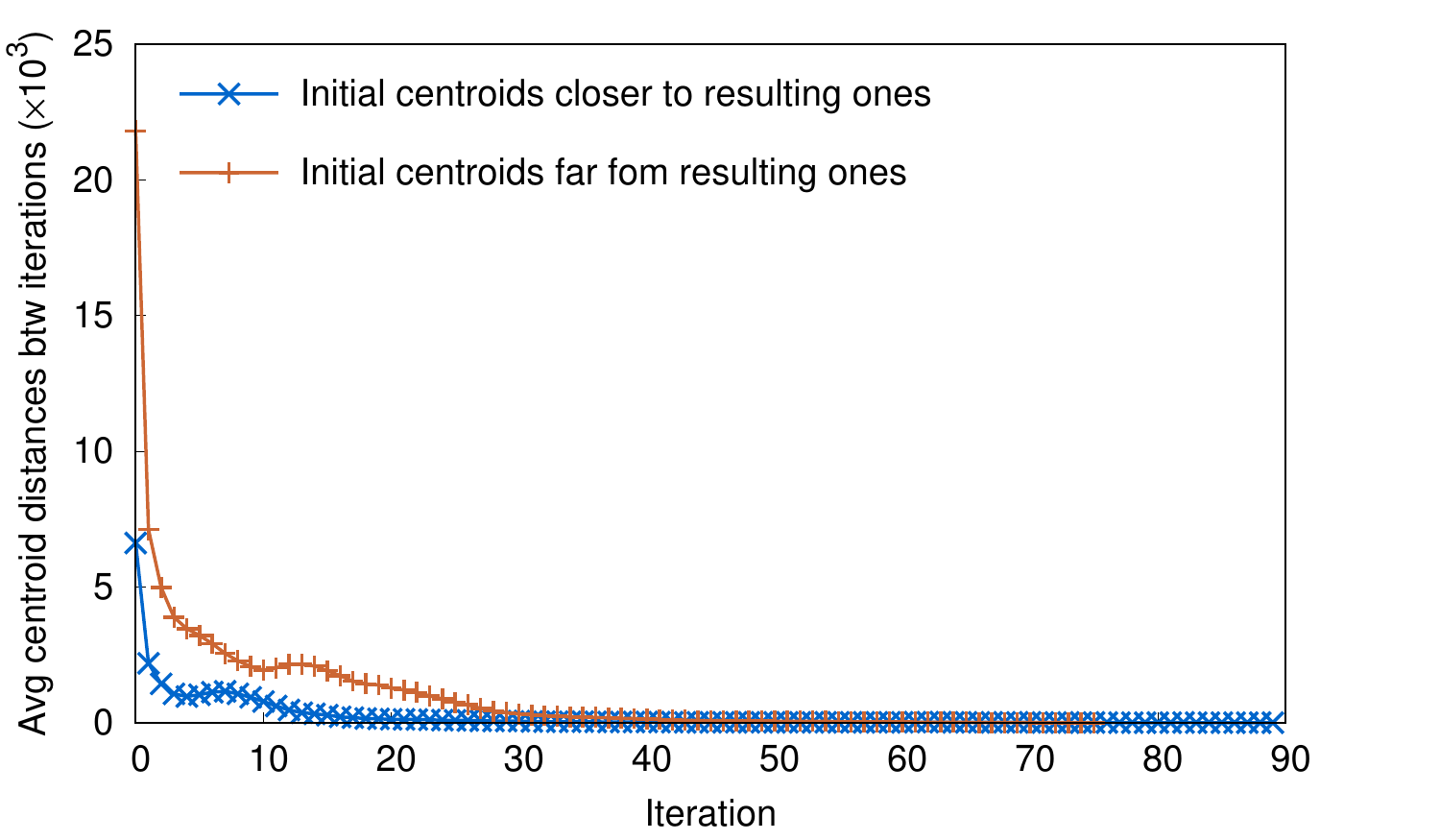}
\caption{\label{fig:convergence}K-means convergence. }
\vspace{-10pt}
\end{figure}

\Cref{fig:kmeans} illustrates 4 out of 76 iterations that K-means took to converge when the threshold was set to zero. 
In this example, we randomly generated 100,000 observations and 10 centers.
As it can be seen in the first frame, the initial centroids were all set on a small area in the bottom-left corner by limiting the domain of generated coordinates. 
In spite of this, they quickly assume positions that are closer to the best fit.
After just 6 iterations, the average distance of centroids decreases more than six times. 
Besides, even with an initial state closer to the final one, we can get a slower convergence (90 iterations, as shown in \Cref{fig:convergence}).
The final result depends on the initial points and there is no guarantee that it is the global optimum, neither about how fast it will converge. 

For the following executions, we arbitrarily set the threshold to be one thousandth of the diagonal of the rectangle that contains all the observed points. 
That means that the iterative process finishes when the average distance of centroids between two subsequent iterations is less than that fraction, so that we can avoid slow convergences. For the given examples of \Cref{fig:convergence}, this criteria would stop the executions by the 41st (instead of 76th) and 21st (instead of 90th) iterations.
From the figure, we can see that there is no much variation from those points on. 

\begin{figure}
\centering
\includegraphics[width=0.5\textwidth]{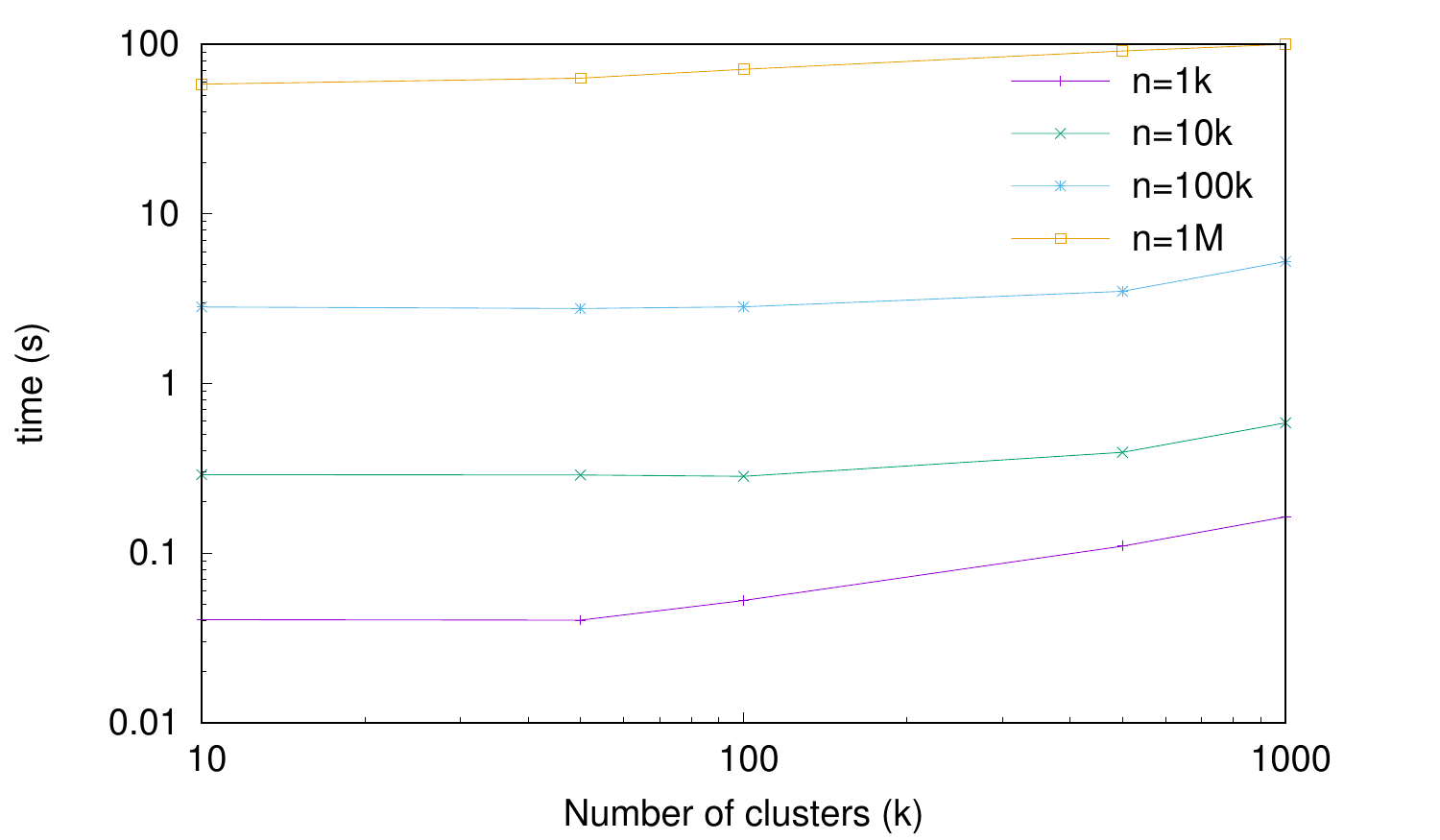}
\caption{\label{fig:varinput}Average time to run one iteration with varying input sizes}
\end{figure}

To assess the influence of input data size, namely the number of observations and centroids, on the memory usage and time consumption, we conducted multiple experiments using the two SGX capable machines described above. 
\Cref{fig:varinput} shows the average time it took to complete one iteration of kmeans with varying input sizes.
It can be noticed that, although the variation on the number of clusters can cause some inflection in the curves, the completion time is mostly affected by the number of observed data points.
Moreover, while the two first increments on the number of data points ($n=10k$ and $n=100k$) caused a proportional increase on consumed time with regards to the data growth (ten times), the last one ($n=1M$) induced a twenty-fold rise.
That can be explained by the growth in the occurrence of cache misses within each worker. 
When that happens, data must be fetched from main memory.
When using SGX protected executions, this means one page has to be evicted from cache (and hence, encrypted), while the one that is fetched must be decrypted and checked for integrity and freshness (that prevents tamper and replay attacks, respectively). 

\begin{figure}
\centering
\includegraphics[width=0.5\textwidth]{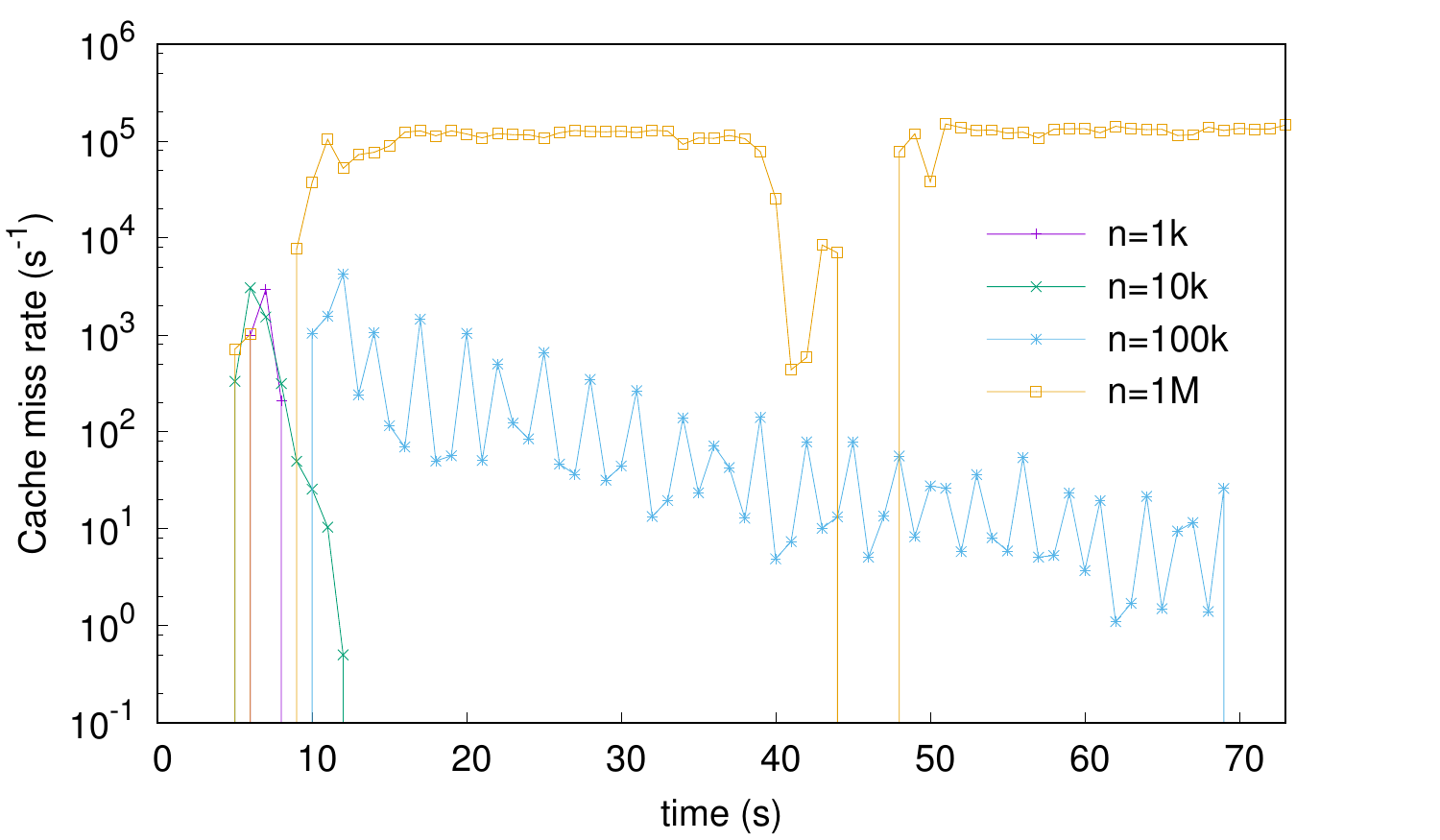}
\caption{\label{fig:cachemisses}Cache miss rates for different input sizes}
\vspace{-10pt}
\end{figure}

The described cache effect is well documented~\cite{Pires:2016,Brenner:2016,Arnautov:2016} and can be noticed by the cache miss rates. 
Since the \emph{reduce} phase is more memory intensive, we chose to make the average on the cache miss rates per second of all 10 reducers in each execution, i.e., for the same second, cache miss rates of reducers were summed and divided by 10. 
Wall clocks were synchronized with a common NTP server, so that the resulting skew was at the range of tens of milliseconds and it should not affect the sampling resolution of 1s.
\Cref{fig:cachemisses} shows that measurement as reported by the tool \texttt{pidstat} when the number of centroids is $k=50$. 
Note that \emph{y} scale is logarithmic, so that the cache miss rates for $n=1M$ is at least two orders of magnitude higher than $n=100k$.

\begin{figure}
\centering
\includegraphics[width=0.5\textwidth]{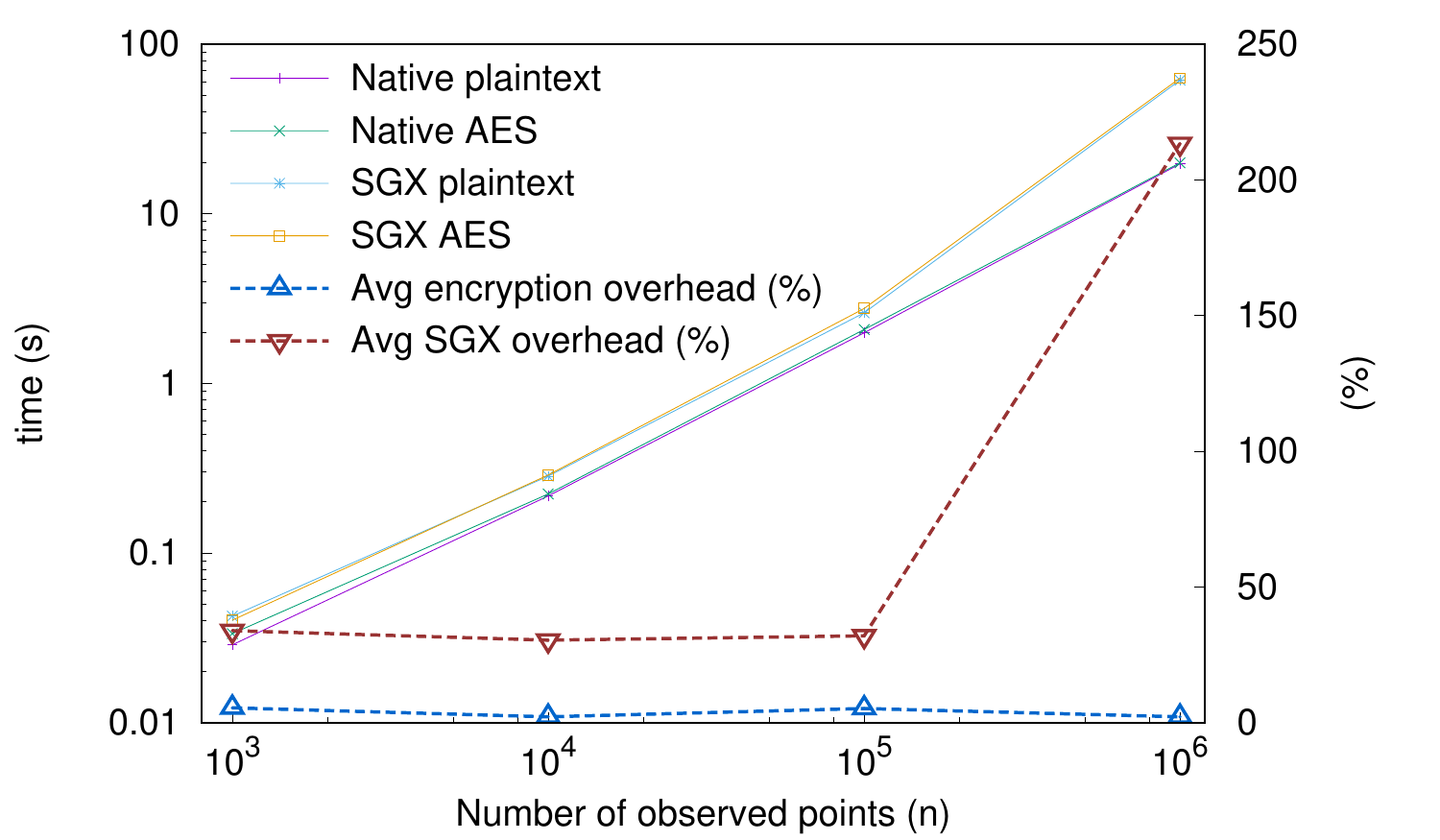}
\caption{\label{fig:sgxandenc}Encryption and SGX overhead}
\end{figure}

Finally, we moved to appraising the influence of SGX and encryption when compared to native executions (no hardware protection).
We ran the same datasets with a fixed number of clusters $k=50$ and observed points varying from $n=1000$ until $n=1M$ along with the four combinations of turning on and off enclaves and encryption.
Results are plotted in \Cref{fig:sgxandenc}. 
We also included the overhead in percentage. 
Encryption overhead was obtained by comparing the average time it took to complete one iteration with and without encryption both inside and outside the enclave, and averaging the two values. 
SGX overhead was established analogously.
The time corresponds to the average of all iterations in a single run of k-means (until the threshold was reached). Coefficient of variation across multiple runs was negligible.
We can notice that encryption overhead is quite low, of around 5\%. 
Enclave execution, on the other hand, is kept around 30\% until we start to get high occurrence of cache misses, as discussed before, when it reaches more than 200\%.

The data volume exchanged in each MapReduce step can be seen in \Cref{tab:datasize}.
Numbers correspond to the average per iteration for the same executions as \Cref{fig:sgxandenc}.
We observe an increase in overheads when cache misses rate grows.
Therefore, based on the processor's cache size, we could establish the maximum amount of data a single SGX-capable machine would be able to handle before incurring in too much overhead.
In our experiments, we perceived that behavior when processing amounts somewhere in between 11 MB and 96.5MB shared between 2 machines (average of around 54MB, or 27MB per machine).
Therefore, a rough estimation would be to limit those amounts to three times the cache size, or 24MB in our case. 
More scalability could be provided horizontally, with the addition of more machines.

\begin{table}[t!]
    \centering
    \begin{tabular}{|c|c|c|c|}
    \hline
            &Split       &Shuffle     &Output \\
    \hline
$n=1k$      &58.7 KB     &112.1 KB    &4.3 KB \\
    \hline
$n=10k$     &257.2 KB    &1.1 MB      &4.5 KB \\
    \hline
$n=100k$    &2.2 MB      &11 MB       &4.6 KB \\
    \hline
$n=1M$      &19.1 MB     &96.5 MB     &4.6 KB \\
    \hline
    \end{tabular}
    \caption{\label{tab:datasize}Data volume exchanged per iteration}
\vspace{-20pt}
\end{table}

\section{Conclusion}
\label{sec:conclusion}

We have proposed in the current work a lightweight framework for implementing secure MapReduce applications in an untrusted environment.
Our approach does not require any particular knowledge from the programmer of cryptographic mechanisms for implementing the \emph{map} and \emph{reduce} functions.
Also, preserving privacy and integrity is not dependent in any way on the specific characteristics of the functions, simply taking advantage in this purpose of the standard TEE characteristics of SGX enclaves.
Besides these advantages, our solution is easily maintainable relying on a standard Lua interpreter for code execution.

The main objective of the current implementation was to show its viability and to assess its behavior under different conditions. 
We consider as a next step testing our prototype at a larger scale and comparing it with other solutions for securing MapReduce (e.g.,~\cite{Schuster:Security:2015,M2R}).
We are also considering taking in account the shuffle phase security in respect to attacks based on traffic analysis. 
We believe that detection of tampering at network level and denial of service attacks could be easily implemented through well known techniques (e.g. sequential numbers, MACs, nonces), which we also leave for future work.


\section*{Acknowledgments}
\begin{wrapfigure}{r}{0.10\textwidth}
\includegraphics[width=\linewidth]{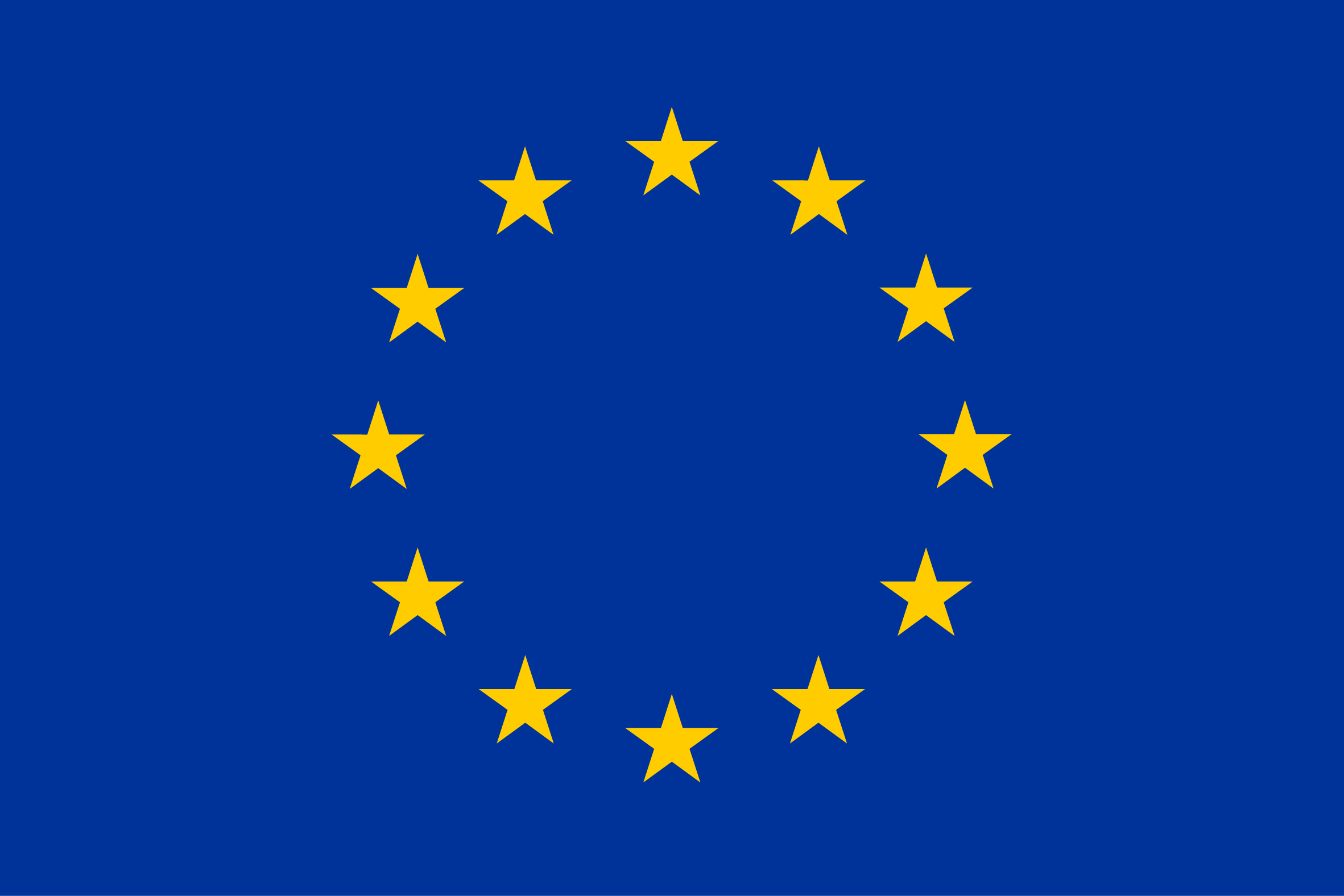} 
\end{wrapfigure}
This work is partly funded from the \emph{European Union{\textquotesingle}s Horizon 2020 research and
innovation programme} under grant agreement No 692178 (EBSIS project) and H2020-ICT-2015 under grant agreement 690111 (SecureCloud project).
This work was partly supported by a grant of the Romanian National Authority for Scientific Research and Innovation, CNCS/CCCDI - UEFISCDI, project number 10/2016, within PNCDI III.
Rafael Pires is also sponsored by CNPq, National Counsel of Technological and Scientific Development, Brazil.



%
%
%

\bibliographystyle{IEEEtran}
\bibliography{references}

\begin{thebibliography}{10}
\providecommand{\url}[1]{#1}
\csname url@samestyle\endcsname
\providecommand{\newblock}{\relax}
\providecommand{\bibinfo}[2]{#2}
\providecommand{\BIBentrySTDinterwordspacing}{\spaceskip=0pt\relax}
\providecommand{\BIBentryALTinterwordstretchfactor}{4}
\providecommand{\BIBentryALTinterwordspacing}{\spaceskip=\fontdimen2\font plus
\BIBentryALTinterwordstretchfactor\fontdimen3\font minus
  \fontdimen4\font\relax}
\providecommand{\BIBforeignlanguage}[2]{{%
\expandafter\ifx\csname l@#1\endcsname\relax
\typeout{** WARNING: IEEEtran.bst: No hyphenation pattern has been}%
\typeout{** loaded for the language `#1'. Using the pattern for}%
\typeout{** the default language instead.}%
\else
\language=\csname l@#1\endcsname
\fi
#2}}
\providecommand{\BIBdecl}{\relax}
\BIBdecl

\bibitem{Dean:2008}
J.~Dean and S.~Ghemawat, ``Map{R}educe: Simplified data processing on large
  clusters,'' \emph{Commun. ACM}, vol.~51, no.~1, pp. 107--113, 2008.

\bibitem{Schuster:Security:2015}
F.~Schuster, M.~Costa, C.~Fournet, C.~Gkantsidis, M.~Peinado, G.~Mainar-Ruiz,
  and M.~Russinovich, ``{VC3}: Trustworthy data analytics in the cloud using
  {SGX},'' in \emph{2015 IEEE Symposium on Security and Privacy}, 2015, pp.
  38--54.

\bibitem{Ohrimenko:2015}
O.~Ohrimenko, M.~Costa, C.~Fournet, C.~Gkantsidis, M.~Kohlweiss, and D.~Sharma,
  ``Observing and preventing leakage in {M}ap{R}educe,'' in \emph{Proceedings
  of the 22Nd ACM SIGSAC Conference on Computer and Communications Security},
  ser. CCS '15.\hskip 1em plus 0.5em minus 0.4em\relax ACM, 2015, pp.
  1570--1581.

\bibitem{M2R}
T.~T.~A. Dinh, P.~Saxena, E.-C. Chang, B.~C. Ooi, and C.~Zhang, ``M2{R}:
  Enabling stronger privacy in {M}ap{R}educe computation,'' in \emph{24th
  USENIX Security Symposium (USENIX Security 15)}.\hskip 1em plus 0.5em minus
  0.4em\relax USENIX Association, 2015, pp. 447--462.

\bibitem{Huang:2010}
S.~Huang, J.~Huang, J.~Dai, T.~Xie, and B.~Huang, ``The {H}i{B}ench benchmark
  suite: Characterization of the {M}ap{R}educe-based data analysis,'' in
  \emph{2010 IEEE 26th International Conference on Data Engineering Workshops
  (ICDEW 2010)}, 2010, pp. 41--51.

\bibitem{Dolev:2016}
S.~Dolev, Y.~Li, and S.~Sharma, ``Private and secure secret shared
  {M}ap{R}educe (extended abstract),'' in \emph{Data and Applications Security
  and Privacy XXX: 30th Annual IFIP WG 11.3 Conference, DBSec 2016, Trento,
  Italy, July 18-20, 2016. Proceedings}, S.~Ranise and V.~Swarup, Eds.\hskip
  1em plus 0.5em minus 0.4em\relax Springer International Publishing, 2016, pp.
  151--160.

\bibitem{Shamir:1979}
A.~Shamir, ``How to share a secret,'' \emph{Commun. ACM}, vol.~22, no.~11, pp.
  612--613, Nov. 1979.

\bibitem{SecureMR}
W.~Wei, J.~Du, T.~Yu, and X.~Gu, ``Secure{MR}: A service integrity assurance
  framework for {M}ap{R}educe,'' in \emph{2009 Annual Computer Security
  Applications Conference}, 2009, pp. 73--82.

\bibitem{Roy:2010}
I.~Roy, S.~T.~V. Setty, A.~Kilzer, V.~Shmatikov, and E.~Witchel, ``Airavat:
  Security and privacy for {M}ap{R}educe,'' in \emph{Proceedings of the 7th
  USENIX Conference on Networked Systems Design and Implementation}, ser.
  NSDI'10.\hskip 1em plus 0.5em minus 0.4em\relax USENIX Association, 2010, pp.
  20--20.

\bibitem{TaggedMR}
C.~Zhang, E.~C. Chang, and R.~H.~C. Yap, ``Tagged-{M}ap{R}educe: A general
  framework for secure computing with mixed-sensitivity data on hybrid
  clouds,'' in \emph{2014 14th IEEE/ACM International Symposium on Cluster,
  Cloud and Grid Computing}, 2014, pp. 31--40.

\bibitem{CS:03}
P.~Eugster, P.~Felber, R.~Guerraoui, and A.-M. Kermarrec, ``The many faces of
  publish/subscribe,'' \emph{ACM Computing Surveys}, vol.~35, no.~2, pp.
  114--131, 2003.

\bibitem{Pires:2016}
R.~Pires, M.~Pasin, P.~Felber, and C.~Fetzer, ``Secure content-based routing
  using {I}ntel {S}oftware {G}uard {E}xtensions,'' in \emph{Proceedings of the
  17th International Middleware Conference}, ser. Middleware '16.\hskip 1em
  plus 0.5em minus 0.4em\relax ACM, 2016, pp. 10:1--10:10.

\bibitem{macqueen1967}
J.~MacQueen, ``Some methods for classification and analysis of multivariate
  observations,'' in \emph{Proceedings of the Fifth Berkeley Symposium on
  Mathematical Statistics and Probability, Volume 1: Statistics}.\hskip 1em
  plus 0.5em minus 0.4em\relax Berkeley, Calif., USA: University of California
  Press, 1967, pp. 281--297.

\bibitem{Lua2007}
A.~Hirschi, ``Traveling light, the {L}ua way,'' \emph{IEEE Software}, vol.~24,
  no.~5, pp. 31--38, 2007.

\bibitem{hintjens2013zeromq}
P.~Hintjens, \emph{ZeroMQ: Messaging for Many Applications}.\hskip 1em plus
  0.5em minus 0.4em\relax O'Reilly Media, Inc., 2013.

\bibitem{StreamHub2013}
R.~Barazzutti, P.~Felber, C.~Fetzer, E.~Onica, J.~Pineau, M.~Pasin,
  E.~Rivi{\`{e}}re, and S.~Weigert, ``{StreamHub}: a massively parallel
  architecture for high-performance content-based publish/subscribe,'' in
  \emph{The 7th {ACM} International Conference on Distributed Event-Based
  Systems, {DEBS} '13}, 2013, pp. 63--74.

\bibitem{StreamHub2014}
R.~Barazzutti, T.~Heinze, A.~Martin, E.~Onica, P.~Felber, C.~Fetzer, Z.~Jerzak,
  M.~Pasin, and E.~Rivière, ``Elastic scaling of a high-throughput
  content-based publish/subscribe engine,'' in \emph{2014 IEEE 34th
  International Conference on Distributed Computing Systems}, 2014, pp.
  567--576.

\bibitem{Daemen:2002}
J.~Daemen and V.~Rijmen, \emph{The Design of Rijndael}.\hskip 1em plus 0.5em
  minus 0.4em\relax Springer-Verlag New York, Inc., 2002.

\bibitem{Brenner:2016}
S.~Brenner, C.~Wulf, D.~Goltzsche, N.~Weichbrodt, M.~Lorenz, C.~Fetzer,
  P.~Pietzuch, and R.~Kapitza, ``Secure{K}eeper: Confidential {Z}oo{K}eeper
  using {I}ntel {SGX},'' in \emph{Proceedings of the 17th International
  Middleware Conference}, ser. Middleware '16.\hskip 1em plus 0.5em minus
  0.4em\relax ACM, 2016, pp. 14:1--14:13.

\bibitem{Arnautov:2016}
S.~Arnautov, B.~Trach, F.~Gregor, T.~Knauth, A.~Martin, C.~Priebe, J.~Lind,
  D.~Muthukumaran, D.~O{\textquoteright}Keeffe, M.~L. Stillwell, D.~Goltzsche,
  D.~Eyers, R.~Kapitza, P.~Pietzuch, and C.~Fetzer, ``{SCONE}: Secure {L}inux
  containers with {I}ntel {SGX},'' in \emph{12th USENIX Symposium on Operating
  Systems Design and Implementation (OSDI 16)}.\hskip 1em plus 0.5em minus
  0.4em\relax USENIX Association, 2016, pp. 689--703.

\end{thebibliography}

\end{document}